\documentclass{appolb}
\usepackage{epsfig}
\usepackage{amsmath}
\usepackage{color}
\usepackage[utf8]{inputenc}

\begin{document}
\title{Study of Tetraquarks with a simple model%
\thanks{Presented at Excited QCD 2011, 20-25 February 2011- Les Houches (France)}%
}
\author{ Marco Cardoso,
Pedro Bicudo,
Nuno Cardoso
\address{CFTP, Instituto Superior Técnico}
}
\maketitle
\begin{abstract}
Tetraquarks are studied with an approximation which reduces the internal degrees of freedom of the system,
using both finite differences and scattering theory. The existence of bound states and resonances is inspected
and the masses and widths of the found resonances are calculated.
\end{abstract}
\PACS{12.38Gc,12.39Mk,12.39Pn,14.40Rt}

\section{Introduction}

For the $ q_1 q_2 \bar q_3 \bar q_4$ system, there is evidence in Lattice QCD
\cite{Okiharu:2004ve,Alexandrou:2004ak},
at least for static quarks, that the hamiltonian, is
\begin{equation}
	H = \sum_{i=1}^4 T_i + V_{1234}
\label{hamiltonian44q}
\end{equation}
with the potential being given by $V_{1234} = \min( V^{Tetra}, V^{M_1 M_2}, V^{M_3 M_4} )$,
where $V^{M_1 M_2}$ and $V_{M_3 M_4}$ are the two possible meson-meson potentials and
\begin{equation}
V^{Tetra} = C + \alpha_s \sum_{i < j} \frac{\lambda_i}{2} \cdot \frac{\lambda_j}{2} \frac{1}{r_{ij}}
	+ \sigma L_{min}( \mathbf{r}_1, \mathbf{r}_2, \mathbf{r}_3, \mathbf{r}_4 )
\end{equation}
is the tetraquark sector potential. We call $V_{1234}$ the triple flip-flop potential.

\begin{figure}[t!]
\begin{center}
\includegraphics[width=0.25\linewidth]{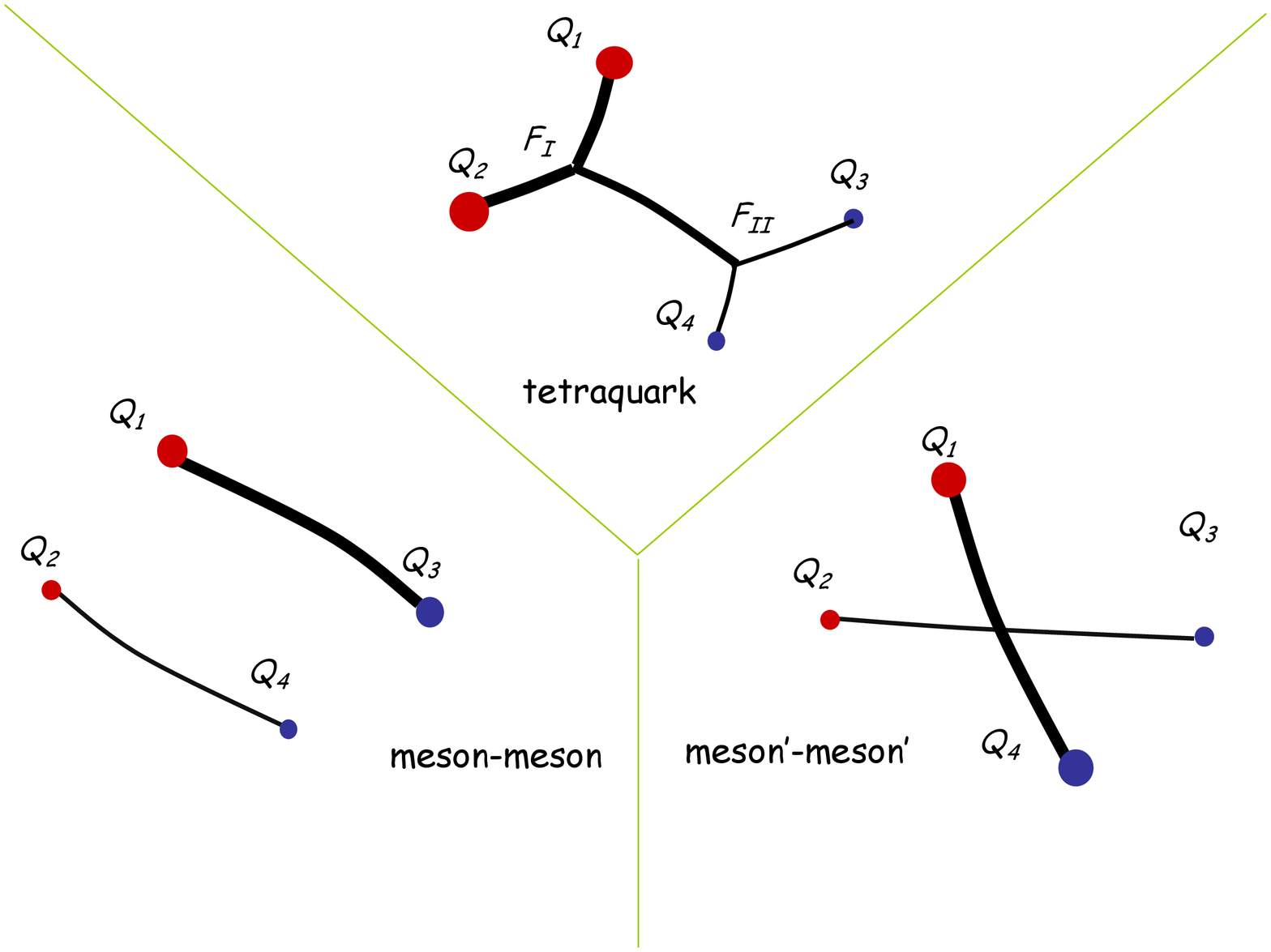}
\includegraphics[width=0.25\linewidth]{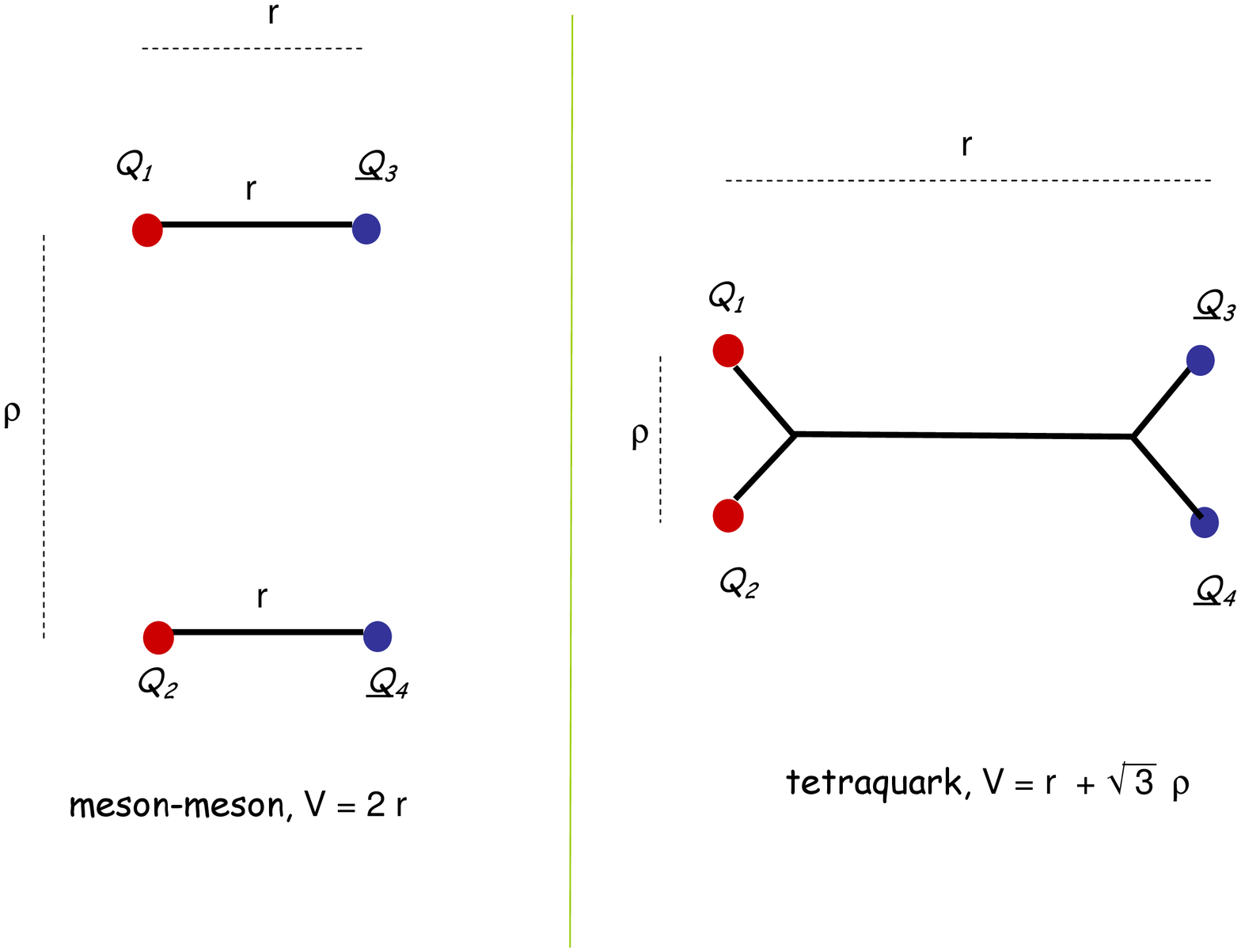}
 \caption{ Left: Triple flip-flop Potential potential.
Right: Simplified potential, which depends only on $\boldsymbol{\rho}$ and $\mathbf{r}$.
\label{potentials}
}
\end{center}
\end{figure}

\begin{figure}[t!]
\begin{center}
\includegraphics[width=0.2\linewidth]{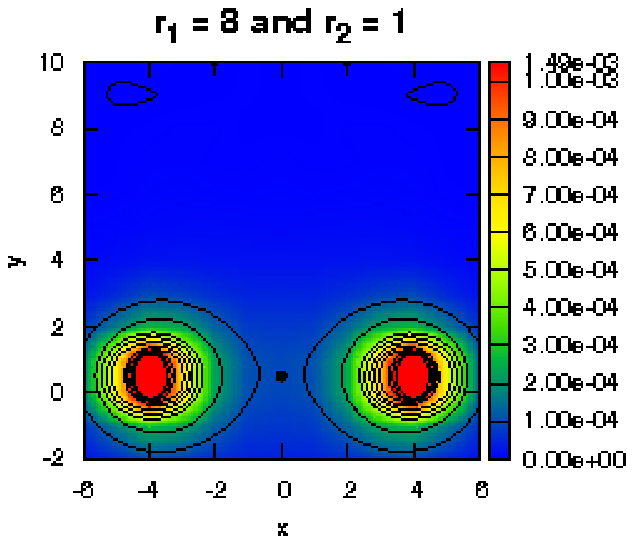}
\includegraphics[width=0.2\linewidth]{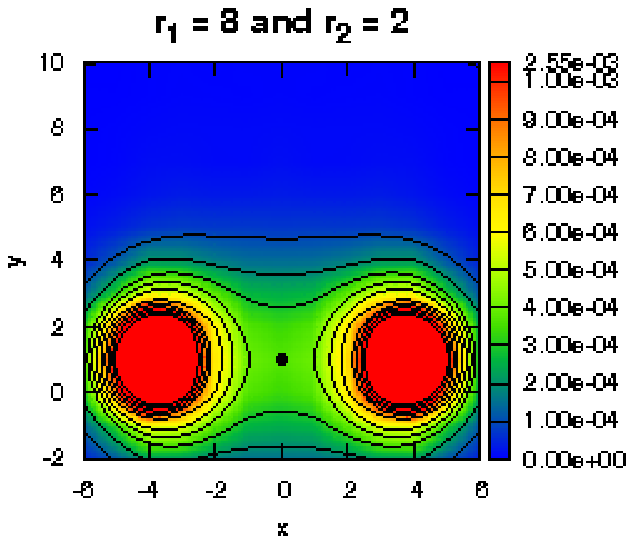}
\includegraphics[width=0.2\linewidth]{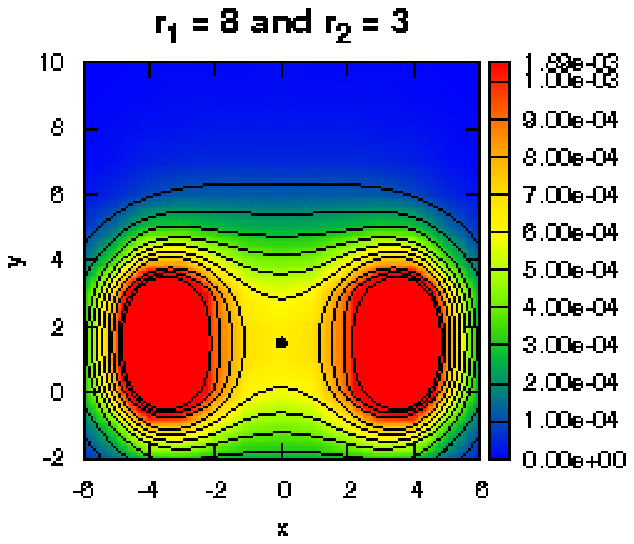}
\includegraphics[width=0.2\linewidth]{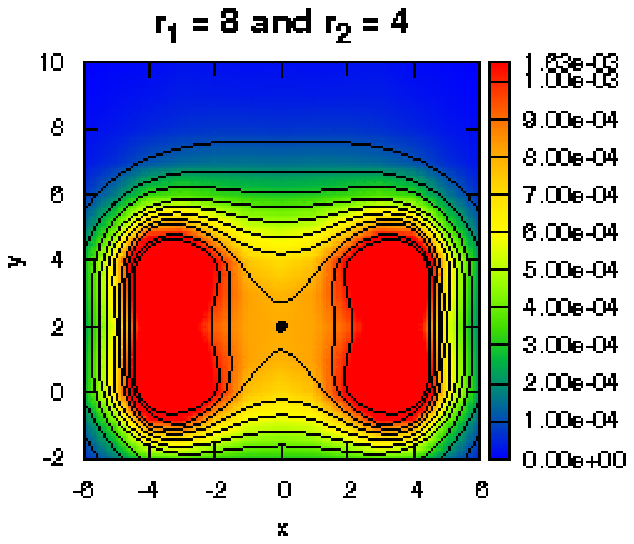}
\includegraphics[width=0.2\linewidth]{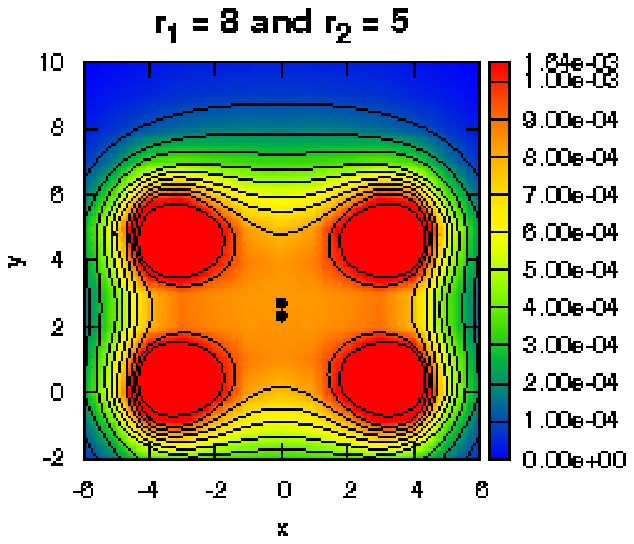}
\includegraphics[width=0.2\linewidth]{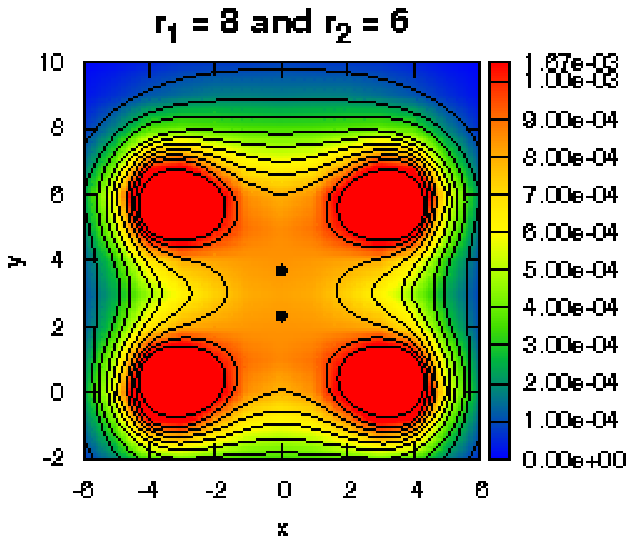}
\includegraphics[width=0.2\linewidth]{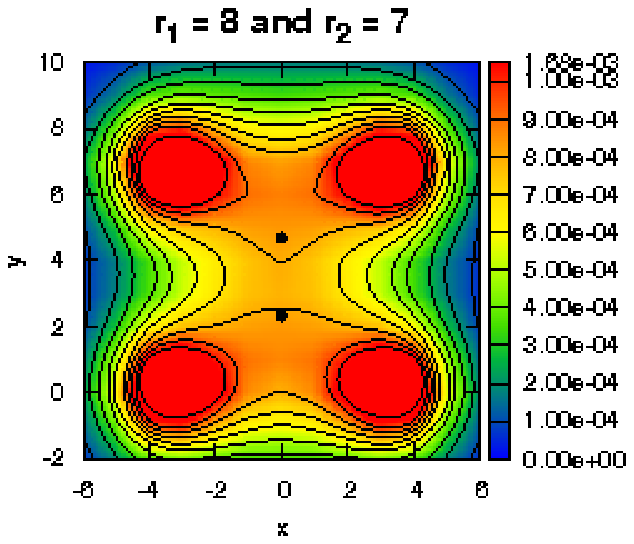}
\includegraphics[width=0.2\linewidth]{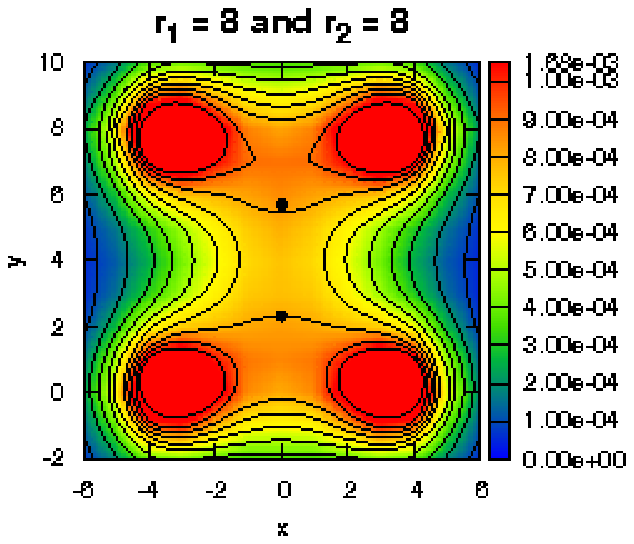}
 \caption{ Plot of Lagrangian density of a static tetraquark. $r_1$ is the intra-diquark distance
and $r_2$ the inter-diquark distance.
\label{chromofields}
}
\end{center}
\end{figure}

In the tetraquark configuration, a single connected string is formed. This string is formed from
five segments of string as can be seen in Fig. \ref{potentials} (see \cite{Bicudo:2008yr,Ay:2009zp}
for a detailed explanation ).
We have studied the colour field configurations for the tetraquark.
The results are shown on Fig. \ref{chromofields}.

It is hard to solve this hamiltonian, so we will instead solve a simplified potential model
which contains the essential ingredients of the triple flip-flop potential, but depends on a smaller number
of variables.

We will do the following approximations. First we will consider that all the quarks and antiquarks
have an equal mass, while neglecting the exchange effects. Secondly, we will consider only the confining (linear)
part of the potential and that the quarks are non-relativistic.
Finally, we will also reduce the number of variables of the model.

\section{A toy model for the potential}

We now simplify the triple flip-flop potential, in order to reduce it to a 
two-dimension problem, similar to the one of a cherry in a broken glass.
To simplify the triple flip-flop potential, we assume that the quark-quark distance
$\boldsymbol{\rho}_{12} = \mathbf{r}_1 - \mathbf{r}_2$ and the antiquark-antiquark distance
$\boldsymbol{\rho}_{34} = \mathbf{r}_3 - \mathbf{r}_4$ are equal
$\boldsymbol{\rho}_{12} = \boldsymbol{\rho}_{34} = \boldsymbol{\rho}$.
Thus, we only need to consider the variable $\boldsymbol{\rho}$ and
$\mathbf{r} = \frac{\mathbf{r}_1 + \mathbf{r}_2}{2} - \frac{\mathbf{r}_3 + \mathbf{r}_4}{2}$
in our model.

In this way, only one meson-meson branch is relevant, and the simplified flip-flop potential is given by
$ V_{FF}(r, \rho) =\min(V_{MM} , V_T) $, where $V_{MM} (r, \rho) = \sigma (2 r) $ and 
$V_T (r, \rho) = \sigma (r + \sqrt 3 \rho)$
This potential is depicted in the right side of Fig. \ref{potentials}.

In this way, we obtain the Schrödinger equation, for this model:
\begin{equation}
\left[ 
- {\hbar^2 \over 2 m} 
\left( \nabla_r^2 + \nabla_\rho^2 \right)
+ 
V_{FF}(r, \rho) 
\right] 
\psi(r, \rho) = E \psi(r , \rho) \, .
\label{schro}
\end{equation}
and what we call the outgoing spherical wave method (useful to study resonances).

\section{Finite difference method}

Since the Hamiltonian, is invariant for independent rotations of $\boldsymbol{\rho}$ and $\mathbf{r}$,
we have two conserved angular momenta $\mathbf{L}_r = \mathbf{r} \times \mathbf{p}_r$ and
$\boldsymbol{L}_{\rho} = \boldsymbol{\rho} \times \mathbf{p}_\rho$.
So, the wavefunction can be written as
$\Psi = \frac{u(r,\rho)}{r \rho} Y_{l_r m_r}(\theta_r, \varphi_r) Y_{l_\rho m_\rho}(\theta_\rho , \varphi_\rho)$
and the Schrödinger equation(\ref{schro})  becomes
\begin{equation}
- \frac{\hbar^2}{2 m} \Big( \frac{d^2}{d r^2} + \frac{d^2}{d \rho^2} \Big) u
	+ \frac{l_r (l_r + 1)}{r^2} u + \frac{l_\rho (l_\rho + 1)}{\rho^2} u + V_{FF}(r,\rho) u = E u \, .
\end{equation}

We now place the system in a lattice and discretize the second derivatives the usual way.
Dirichlet boundary conditions are used at the origin, for the wavefunction to be regular
and at the box boundary for convenience.

We then evaluate the eigenstates of the system. Only a few of them are bound or semi-localized states.
They can be identified by calculating $\sqrt{\langle \rho^2 \rangle}$ and comparing it to the box size.

\begin{figure}[t!]
\begin{center} 
  \includegraphics[width=0.25\linewidth]{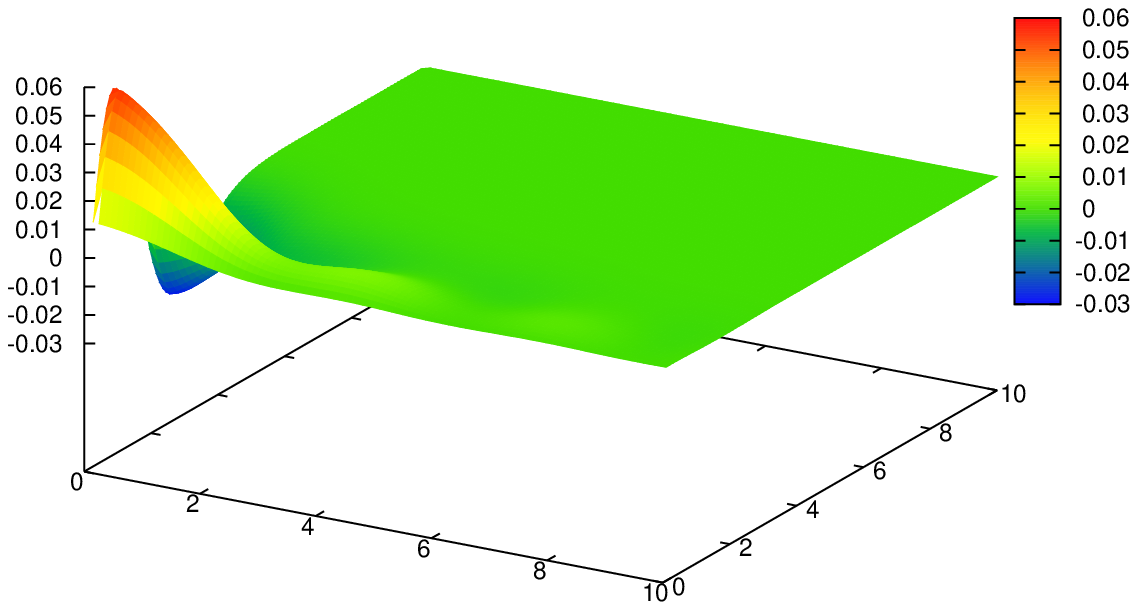}
  \includegraphics[width=0.25\linewidth]{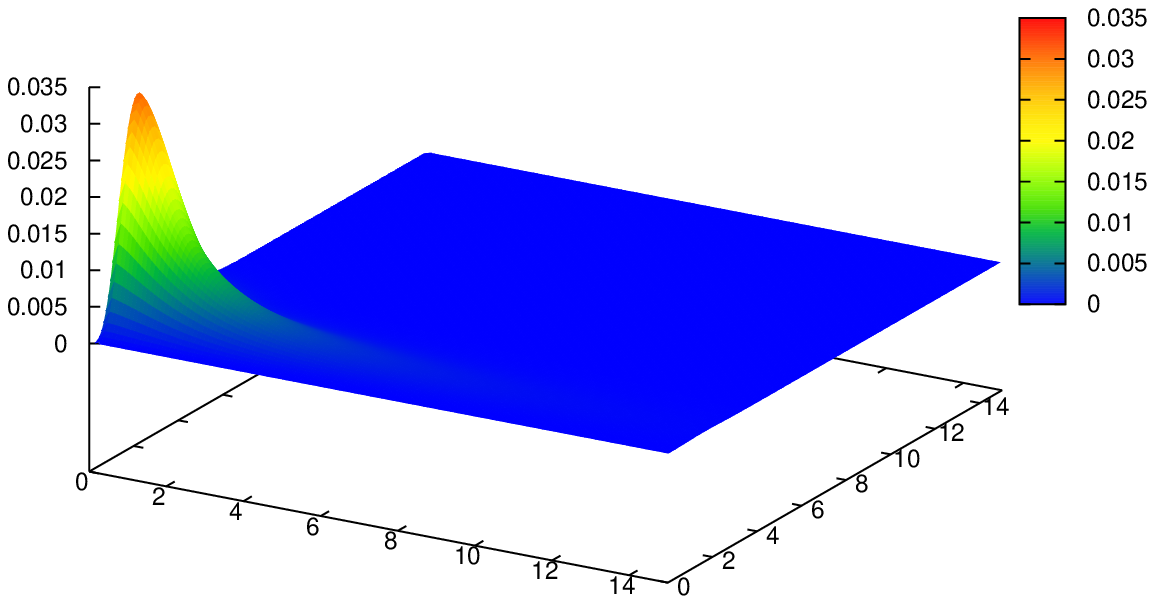}
  \includegraphics[width=0.25\linewidth]{delta.eps}
\caption{Left: Semi-localized state, or resonance for $l_r = 1$.
  Center: Bound state for $l_r = 3$.
  Right: "Phase shifts" obtained from the finite differences.
\label{finitediff_res}
}
\end{center}
\end{figure}

In the following results, we consider only $l_\rho = 0$. 
For $l_r = 0$, only continuum states are found. However, for $l_r > 0$, we are able to found some semi-localized
states, such as the one depicted on the left side of Fig. \ref{finitediff_res}.
On the center, we can see a true bound state found for $l_r = 3$.

We try to calculate the phase shifts using the finite differences results, by calculating the functions
$\psi_i(\boldsymbol{\rho}) = \int d^3 \mathbf{r} \phi_i^*(\mathbf{r}) \, \Psi(\mathbf{r},\boldsymbol{\rho})$
where the $\phi_i$ are the eigenfunctions of the hamiltonian
$\hat{H}_{MM} = - \frac{\hbar^2}{2 m} \nabla_r^2 + 2 \sigma r$
and, then by fitting the long range limit $\psi_i \rightarrow A_i \sin( k_i \rho + \delta_i )$.
We obtain the results in the right side of Fig. \ref{finitediff_res}.
As we can see there, the results become irregular when a second channel is open, due the interference
of the different channels, and so we can't use this method to calculate the phase shifts.

For that, we had to resort to a scattering theory approach.

\section{Outgoing spherical wave method}

By expanding the wavefunction as $\Psi(\boldsymbol{\rho},\mathbf{r}) = \sum_i \phi_i(\mathbf{r},\boldsymbol{\rho})$
we obtain the coupled channel Schrödinger equation
\begin{equation}
- \frac{\hbar^2}{2 \mu_i} \nabla^2 \Psi_{i} + V_{ij} \Psi_j = ( E - \epsilon_i ) \Psi_i \ ,
\label{schro_cc}
\end{equation}
where $V_{ij} = \int d^3 \mathbf{r} \phi_i*(\mathbf{r}) ( V_{FF} - 2 \sigma r ) \phi_j(\mathbf{r})$,
$\epsilon_i$ are the eigenvalues of $\hat{H}_{MM}$ and $\mu_i = m$.

Considering the scattering from the channel $i$ to the channel $j$, 
we have asymptotically $\Psi_j \rightarrow \delta_{ij} e^{i k_i z} + f_{ij}( \hat{r} ) \frac{e^{ i k_j r } }{ r }$,
if the channel $j$ is open, otherwise vanishing.
The $f_{ij}$ can be computed by considering the outgoing solutions of the Eq. (\ref{schro_cc}) ,
$\Psi_j = e^{i k_i r} \delta_{ij} + \sum_{nm} \frac{u_n^l(r)}{r} Y_{l m}(\theta,\varphi)$.
We obtain \cite{Bicudo:2010mv}
\begin{equation}
- \frac{\hbar^2}{2 \mu_j} \frac{d^2 u_j}{dr^2} + V_{jn} u_n = ( E - \epsilon_j ) u_j - V_{jn} j_l( k_i r ) r \ .
\label{theeq}
\end{equation}
Also we can obtain the phase shifts by studying the asymptotic behaviour of the solutions.

We can describe the scattering process with four quantum numbers: The scattering angular momentum $l_\rho$, the confined
angular momentum $l_r$ and the initial and final states radial number in 
the confined coordinate $r$, $n_i$ and $n_j$.

We now compute the phase shifts, in order to search for resonances in our simplified flip-flop model.

The energies in the plots will be given in terms of $E_0 = (\hbar^2 \sigma^2 / m)^{1/3}$, which is
the energy scale of the problem.

\begin{figure}[t!]
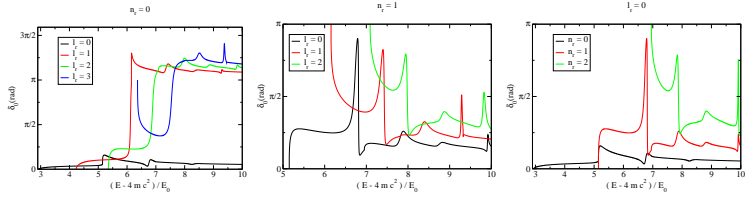

\begin{center}
  \includegraphics[width=0.25\linewidth]{delta_c0_lrho0_E.eps}
  \includegraphics[width=0.25\linewidth]{delta_c1_lrho0_E.eps}
  \includegraphics[width=0.25\linewidth]{delta_lr0_lrho0_E.eps}
\caption{ Comparison of the phase shifts for $l_r = 0, 1, 2$ and $3$, with $n_r = 0$ (left) and $n_r = 1$ (center).
 In the right, phase shifts, for $l_r = 0$ and different $n_r$.
\label{deltas}
}
\end{center}
\end{figure}

In Fig. \ref{deltas} we see the phase shifts for different values of $l_r$, from the radial ground state
$n_r = 0$.
As can be seen, no resonance is formed for $l_r = 0$, but for $l_r = 1$ and $l_r = 2$ one resonance is formed
since the phase shift has a jump of $\pi$. This is somewhat expected, since the centrifugal barrier has the effect
of maintaining the diquarks separated, while increasing the stability of the system.
We also see a bound state for $l_r = 3$, which agrees with the results obtained with the finite
differences method and presented in Fig. \ref{finitediff_res}.
The phase shifts for scattering for $n_r = 1$ and different $l_r$ and for $l_r = 0$ and different values
of $n_r$ are also presented.

\begin{figure}[t!]
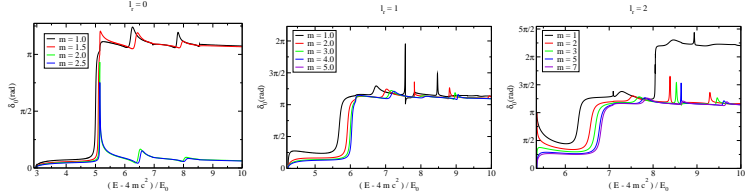

\begin{center}
  \includegraphics[width=0.25\linewidth]{redmass_lr0_E.eps}
  \includegraphics[width=0.25\linewidth]{redmass_lr1_E.eps}
  \includegraphics[width=0.25\linewidth]{redmass_lr2_E.eps}
\caption{ Phase shifts, including the reduced mass correction for $l_r = 0$, $1$ and $2$.
\label{redmass}
}
\end{center}
\end{figure}

If we include the relativistic corrections for the reduced mass of the system, the $\mu_i$ becomes
$\mu_i = m + \frac{\epsilon_i}{4 c^2}$.
Greater reduced masses could increase the binding of the system.
Which is in fact what we observe in Fig. \ref{redmass}.
For $l_r = 0$ we can see a resonance that is destroyed for masses as low as $2 \sqrt{\sigma} / c^2$, while
for $l_r = 1$ the resonance is always present.
When $l_r = 2$ for sufficiently low masses we can observe a bound state and even a resonance in addition to it.

Finally we can compute the decay widths of the resonances
by using the expression
${\Gamma \over 2} = \left( d \delta \over d E \right)^{-1}$
when $\delta$ is $\pi/2$. 
The results are given in Table \ref{widthtables}.
The width is larger for large angular momentum and small quark mass.

Using this results to estimate the physical scale of our decay widths, we obtain, in the $l_r = 1$ state,
a decay width $\Gamma_l \simeq 15 MeV$ for light quarks ($m \simeq 0.4$ GeV).
For the charm quark we obtain $\Gamma_c \simeq 10 \mbox{MeV}$ and for the bottom quark, 
the estimated width is $\Gamma_b \simeq 6 \mbox{MeV}$.

In the $l_r = 2$ state the estimated widths are $\Gamma_l = 52 \mbox{MeV}$, $\Gamma_c = 34 \mbox{MeV}$
and $\Gamma_b = 23 \mbox{MeV}$, while in the $l_r = 3$ state, the values are
$\Gamma_l = 140 \mbox{MeV}$, $\Gamma_c = 90 \mbox{MeV}$ and $\Gamma_b = 60 \mbox{MeV}$.

\begin{table}
\begin{centering}
\begin{tabular}{|c|c|c|}
	\hline
	$l_r$ & $(E - 4 m c^2)/E_0$ & $\Gamma$ / $E_0$ \\
	\hline
	1 & 6.116 & 0.037 \\
	2 & 6.855 & 0.131 \\
	3 & 7.462 & 0.352 \\
	\hline
\end{tabular}
\begin{tabular}{|c|c|c|c|}
	\hline
	$l_r$ & $m / \sqrt{\sigma}$ &  $(E - 4 m c^2)/E_0$ & $\Gamma$ / $E_0$ \\
	\hline
	0 & 1.0 & 5.001 & 0.039 \\
	 & 1.5 & 5.096 & 0.022 \\
	1 & 1.0 & 5.659 & 0.137 \\
	 & 3.0 & 5.990 & 0.075 \\
	 & 5.0 & 6.053 & 0.053 \\
	2 & 1.0 & 6.194 & 0.586 \\
	 & 3.0 & 6.634 & 0.209 \\
	 & 7.0 & 6.777 & 0.162 \\
	\hline
\end{tabular}
\caption{Left: Decay widths as a function of $l_r$. Right: Decay widths with the reduced mass correction.}
\label{widthtables}
\end{centering}
\end{table}

\section{Conclusion and outlook}

We have constructed a simplified model for the tetraquark. With this model, we have observed the formation
of a resonance for non null $l_r$, and even a bound state for $l_r = 3$.
So, we conclude that a higher angular momentum favours the formation of resonances and bound states, in
agreement with Karliner and Lipkin \cite{Karliner:2003dt}.
Extrapolating this results to the real world, it seems plausible the existence of tetraquarks with high angular
momentum.

The formation of resonances and bound states is also favoured by low quark masses in our model.
This happens since a low quark mass produces a higher binding energy which increasing the meson masses and
so favouring the localization of resonances and bound states.

We think the methods developed here are useful and could be futurely used in the complete tetraquark problem
with a triple flip-flop potential.


We thank George Rupp for useful discussions. We acknowledge the 
financial support of the FCT grants CFTP, CERN/FP/109327/2009 and
CERN/FP/109307/2009.



\begin{thebibliography}{00}

\bibitem{Okiharu:2004ve}
  F.~Okiharu, H.~Suganuma and T.~T.~Takahashi,
  Phys.\ Rev.\  D {\bf 72}, 014505 (2005)
  [arXiv:hep-lat/0412012].

\bibitem{Alexandrou:2004ak}
  C.~Alexandrou, G.~Koutsou,
  Phys.\ Rev.\  {\bf D71}, 014504 (2005).
  [hep-lat/0407005].

\bibitem{Bicudo:2008yr}
  P.~Bicudo and M.~Cardoso,
  Phys.\ Lett.\  B {\bf 674}, 98 (2009)
  [arXiv:0812.0777 [physics.comp-ph]].

\bibitem{Ay:2009zp}
  C.~Ay, J.~-M.~Richard, J.~H.~Rubinstein,
  Phys.\ Lett.\  {\bf B674}, 227-231 (2009).
  [arXiv:0901.3022 [math-ph]].

\bibitem{Bicudo:2010mv}
  P.~Bicudo, M.~Cardoso,
  Phys.\ Rev.\ {\bf D83}, 094010 (2011)
  [arXiv:1010.0281 [hep-ph]].

\bibitem{Karliner:2003dt}
 M.~Karliner and H.~J.~Lipkin,
 Phys.\ Lett.\  B {\bf 575}, 249 (2003)
 [arXiv:hep-ph/0402260].

\end{thebibliography}
\end{document}